# Charged domain walls and crystallographic microstructures in hybrid improper ferroelectric $Ca_{3-x}Sr_xTi_2O_7$

Hiroshi Nakajima[1], Koji Shigematsu[2], Yoichi Horibe[3], Shigeo Mori[4], and Yasukazu Murakami[1,2]

[1] *Department of Applied Quantum Physics and Nuclear Engineering, Kyushu University, Fukuoka 819-0395, Japan.*

[2] *The Ultramicroscopy Research Center, Kyushu University, Fukuoka, 819-0395, Japan.*

[3] *Department of Materials Science and Engineering, Kyushu Institute of Technology, Kitakyushu 804-8550, Japan.*

[4] *Department of Materials Science, Osaka Prefecture University, Sakai, Osaka 599-8531, Japan.*

The charged domain walls in ferroelectric materials exhibit intriguing physical properties. We examine herein the charged-domain-wall structures in $Ca_{3-x}Sr_xTi_2O_7$ using transmission electron microscopy. When viewed along the [001] axis, the wavy charged domain walls are observed over a wide range (>5 μm). In contrast, short charged-domain-wall fragments (from 10 to 200 nm long) occur because they are intercepted and truncated by the conventional 180° domain walls. These results reveal the unusual charged domain structures in $Ca_{3-x}Sr_xTi_2O_7$ and will be useful for understanding their formation process.

## 1. Introduction

Ferroelectric materials undergo spontaneous electric polarization induced by the displacements of ions, which breaks the charge neutrality in a crystal. To minimize the electrostatic energy, these materials usually develop 90° or 180° ferroelectric domains, both of which can be free from the bound charge. However, some ferroelectric materials develop unusual parallel (head-to-head) and antiparallel (tail-to-tail) polarization alignments between the neighboring domains. In this type of ferroelectric configuration, the domain walls are subjected to bound charge.[1,2]

Recently, Oh *et al*. reported that charged domain walls (CDWs) are produced in the improper hybrid ferroelectric $Ca_{3-x}Sr_xTi_2O_7$[3]. In an improper ferroelectric, electric polarization is the secondary, not primary, order parameter while in a proper ferroelectric, such as $BaTiO_3$, electric polarization is the primary order parameter. The polarization in improper ferroelectrics may be induced by a change in the structural order parameter (i.e., the primary order parameter in this case), including the lattice distortion. Ferroelectricity in $Ca_{3-x}Sr_xTi_2O_7$ has been observed both by conventional electric polarization measurements and by piezoresponse force microscopy.[3] The latter revealed two types of CDWs in $Ca_{3-x}Sr_xTi_2O_7$ in terms of polarity and conductivity; namely, "high-conductivity head-to-head walls" and "low-conductivity tail-to-tail walls." From an engineering viewpoint, CDWs can be promising for device applications because their morphology and conductivity can be controlled by electric fields and other physical phenomena.[4–9] Because microscopy studies on CDWs are vital for improving our understanding of the physical properties of $Ca_{3-x}Sr_xTi_2O_7$, we report herein a transmission electron microscopy (TEM) study on the CDWs in this material.

The domain structures in $Ca_{3-x}Sr_xTi_2O_7$ have already been the subject of several microscopy studies. For example, meandering CDWs were observed to occur in $Ca_{3-x}Sr_xTi_2O_7$ ($x = 0.45$[10] and $0.95$[11]) on the basis of the observations at [001] incidence: i.e., observations from the (001) planes of the crystal. In addition, straight (planar) CDWs occurred in $Ca_{3-x}Sr_xTi_2O_7$ ($0.45$[12] and $0.50$[13]) although these results were obtained at [010] incidence: i.e., observations from the (010) planes of the crystal unlike that in the studies just mentioned[10,11].

The lattice symmetry of $Ca_{3-x}Sr_xTi_2O_7$ depends on the Sr content $x$. Thus, to obtain a thorough understanding of the wall structure, CDWs should be observed from multiple directions in samples consisting of a specific composition of the crystalline $Ca_{3-x}Sr_xTi_2O_7$. In addition, we need to understand the relation between CDWs and the structural imperfections in this crystal [i.e., the antiphase boundary (APB) and irregular stacking of atomic planes], which is of particular importance in materials engineering. Thus, this study uses TEM to investigate the morphology of CDWs in two different crystallographic planes of a single crystal of $Ca_{3-x}Sr_xTi_2O_7$ ($x = 0.45$) and the relation between the CDWs and lattice imperfections. The results provide essential information for understanding the nature of CDWs, which should be of use for future device applications.

## 2. Methods

A single crystal of $Ca_{3-x}Sr_xTi_2O_7$ ($x = 0.45$) (hereinafter CSTO) was grown by using the floating zone method. The single crystal was easily cleaved along the (001) plane, allowing small pieces of the crystal to be prepared. The optical micrograph shown in Fig. 1(a) reveals twin plates in the cleaved (001) surface. For TEM studies, two types of thin-foil specimens showing the normal to [001] and [010] axes were prepared by using a focused-ion-beam system (SEM-FIB, HITACHI MI4000L, acceleration voltage 30 kV). Surface layers that were damaged by the ion beam were removed by argon-ion milling at acceleration voltage 900 V (Fischione Nano Mill Model 1040). The [001]-normal specimen was cut from an area 300 μm away from the [010]-normal one. Note that the crystallographic indices used herein are based on the orthorhombic structure (space group $A2_1am$), in which the electric polarization is parallel to the [100] axis, as illustrated in Fig. 1(b).

Images were acquired by using a transmission electron microscope (JEM-2100HC, JEOL Co. Ltd., acceleration voltage 200 kV). To obtain image contrast to detect the ferroelectric domains, we used dark-field imaging, which is based on the breakdown of Friedel's law[14]. High-angle annular dark-field scanning TEM (HAADF-STEM) images were obtained by using a transmission electron microscope equipped with a spherical-aberration corrector (JEM-ARM200F, JEOL Co. Ltd., acceleration voltage was 200 kV). The probe semi-angle and current were 18.6 mrad and 9 pA, respectively. The angular-detection range of the HAADF detector for scattered electrons was 50–150 mrad. In addition, to gain a deeper understanding of the irregularity in atom stacking (discussed later in detail), we applied off-axis electron holography with an HF3300X (Hitachi Co. Ltd.)

at an acceleration voltage of 300 kV. Double-biprism electron holography was used to remove the Fresnel fringes due to the biprism.[15)] To enhance the signal-to-noise ratio, 120 phase images were averaged.

## 3. Results and discussion

Figure 2(a) shows a dark-field image of CSTO, observed by using reflection 300. The electrons were incident along the [001] axis. Although the reflection 300 is forbidden for the $A2_1am$ lattice, the reflection spot is visible because of multiple scattering [see inset of Fig. 2(a)]. Thus, the spot carries information about the superlattice reflections in the crystal. In addition, this area of the specimen exhibits mosaic contrast. To understand the origin of the mosaic contrast, we collected other dark-field images using different reflections. Figures 2(b) and 2(c) provide dark-field images from the fundamental reflections $\bar{2}00$ and 200, respectively. Note that we used two-beam excitation to acquire these dark-field images [Figs. 2(b) and 2(c)]. The field of view is indicated by the square in Fig. 2(a). Although the left domain is brighter than the right domain in Fig. 2(b), the reverse is true in Fig. 2(c). This contrast reversal may be reasonably understood on the basis of the breakdown of Friedel's law in ferroelectric crystals.[14)]

This image-contrast analysis enables us to determine the polarization direction[14)], with the results shown by the red arrows in Figs. 2(b) and 2(c). For convenience, the position of the ferroelectric domain wall is indicated by the red dotted line in the bright-field image of Fig. 2(e). Note the needle-like contrast in the lower-left of Fig. 2(b) [and 2(c)], which is attributed to other ferroelectric domains that were superimposed in the image projection. Figure 2(d) shows an enlarged view of the square region in Fig. 2(a): (i.e., a dark-field image obtained by using the reflection 300). This result reveals another type of interface that crosses the ferroelectric domain wall, as indicated by the blue dotted line in Fig. 2(e). This interface is attributed to an APB (i.e., a type of planar defect whereby the geometric phase shifts within the crystal) since it is invisible in the dark-field images acquired from the fundamental reflections[16)] [i.e., Figs. 2(b) and 2(c)]. On the basis of the images shown in Fig. 2, we conclude that the CDWs produced in CSTO depend neither on APBs nor on twin boundaries because these interfaces do not correlate with the CDWs.

We now discuss in greater detail the relationship between the ferroelectric domain walls and the APBs. Considering the CSTO crystal, we see that tilting and rotation of the oxygen octahedra result in improper ferroelectricity[10]: Fig. 3 shows the tilting (blue arrows) and rotation (orange arrows) that can occur in CSTO crystals. As shown within the unit cells framed by the solid lines in Figs. 3(a) and 3(b), the sense of rotation (clockwise or counterclockwise) in the unit cell depends on the ferroelectric domain, which leads to polarization reversal. Polarization reversal also occurs at the ferroelectric domain wall between Figs. 3(c) and 3(d). As shown schematically in Figs. 3(a) and 3(c), an APB shifts the geometric phase of the unit cell by half the length of the [110] vector: see the dotted squares (with phase shift) and the solid squares (without phase shift). The APB does not induce polarization reversal, as seen in Figs. 3(a) and 3(c) [and in Figs. 3(b) and 3(d)]. The absence of the polarization reversal is consistent with the observation in Figs. 2(b) and 2(c) in which the APB contrast is negligible.

To compare the CDW structures observed along the [001] axis, we acquired a dark-field image from the specimen that showed the [010] foil normal [Fig. 4(a)]. The image was obtained by using the fundamental 202 reflection [see this reflection in Fig. 4(b)]. Again, the image contrast of the ferroelectric domain walls is attributed to the breakdown of Friedel's law. The orientation of the electric polarization is indicated by the red arrows. Unlike the results shown in Fig. 2, in which the ferroelectric domains are viewed along the [001] axis, most of the domain walls between areas of antiparallel electric polarization are planar. The plates of 180° ferroelectric domains are elongated in the [100] direction. The plate width ranges from 8 to 200 nm depending on the position in the specimen. In addition to the planar 180° domains with no bound charges, both *head-to-head* and *tail-to-tail* CDWs appear in several regions in Fig. 4(a). Interestingly, some of the CDWs terminate at the conventional 180° walls, as indicated by the blue arrowheads. These results indicate that the 180° walls can play an essential role in controlling the morphology and volume fraction of CDWs.

Figure 4(c) summarizes the analysis of the morphology in the three-dimensional structure of the ferroelectric domains: i.e., the combined observations along both the [001] and [010] axes. For [001] incidence, highly meandering CDWs appear over a wide range (>5 μm) with no discontinuities. In contrast, for the [010] incidence, only short fragments of CDWs appear (ranging from 8 to 200 nm in Fig. 4) because they are intercepted by the planar 180° walls. Thus, the plate width of the ferroelectric domains with bound charges depends on the viewing direction.

To obtain further information on the microstructure observed along the [010] axis, we used HAADF-STEM. Figure 5(a) shows a HAADF-STEM image with anomalous contrast (i.e., bright bands), as indicated by the blue and red arrowheads. Figures 5(b)–5(d) show lattice images (HAADF-STEM images) collected from the areas marked "b"–"d" in Fig. 5(a), respectively. The bright dots represent atomic columns. In Fig. 5(b), the two types of atomic columns are indicated by green (pink) for Ca (Sr) and yellow for Ti.

Note that CSTO is a Ruddlesden–Popper-type oxide, $(AO)$–$(A_nB_nO_{3n})$, in which the rock-salt block $AO$ is stacked with the perovskite block $A_nB_nO_{3n}$ [see schematic representation in Fig. 1(b)]. Here, $A$ is an alkaline-earth-metal ion, $B$ is a transition-metal ion, and $n$ is the number of simple perovskite-type unit cells that can be stacked along the [001] axis. The crystal structure schematized in Fig. 1(b) corresponds to $n = 2$ because it contains two perovskite units along the [001] axis. In the image of Fig. 5(b), which was acquired from a defect-free region within a ferroelectric domain, the $n = 2$ perovskite block is made of five-layered stacks of basal planes [i.e., the (001) plane], as shown by the bright dots. The perovskite blocks are intergrown by rock-salt blocks that create the dark lines that separate the groups of five basal planes in Fig. 5(b). Only the $n = 2$ perovskite blocks appear in Fig. 5(b).

Conversely, the images in Figs. 5(c) and 5(d) reflect an irregularity in the stacking number $n$. For example, the HAADF-STEM image in Fig. 5(d) shows an $n = 12$ perovskite block in which a 25-layer stack of basal planes appears without intergrowth of the rock-salt block. On the basis of these HAADF-STEM images, the bright bands in Fig. 5(a) are identified as regions with irregular stacking numbers, such as $n = 3$ (7-layer stack of basal planes), $n = 6$ (13-layer stack of basal planes), $n = 11$ (23-layer stack of basal planes), and $n = 12$. Although Fig. 5 displays perovskite blocks with $n = 3, 6, 11$, and 12, other observation results from this study found perovskite blocks with numbers $n = 5$ and 7. The specimen likely showed other numbers of stacking irregularity (e.g., $n = 4, 8, 9$, and 10). Note that similar irregular regions of perovskite blocks are observed in other Ruddlesden–Popper oxides, such as $Sr_2LnMn_2O_7$ ($Ln$ = Y, La, Eu, or Ho).[17,18]

These irregular regions of perovskite blocks with $n$ = 3, 6, 11, and 12 are likely to be potential sites of 180° ferroelectric domain walls because the wall energy can be minimized at these imperfections. As indicated by the blue arrows in Fig. 1(b), the net polarization in the perovskite blocks decreases with the increasing stacking number $n$ because polarization is the sum of the antiparallel displacements of the Ca and Sr ions per unit volume. Thus, an irregularity in stacking number reduces the ferroelectric order. Actually, a previous study reported that the $n$ = 3 region coincides with a 180° ferroelectric domain wall[13]. Because regions with irregular $n$ can be placed at random in the CSTO crystal, the 180° domains have a wide range of plate width, as shown in Fig. 4. It is thus likely that the terminated CDWs [see blue arrowheads in Fig. 4(a)] may also be understood in terms of the instability of the ferroelectric state in the irregular perovskite blocks.

Electron holography provides another key to better understand the features caused by the irregular stacking number $n$ in CSTO crystal. Electron holography determines the phase shift of electrons that traverse a thin-foil specimen with a mean inner potential $V_0$.[19] The value of $V_0$ depends on the elements and atomic density in the specimen. Assuming electric charging[20] of the foil due to the electron exposure can be ignored, the phase shift $\varphi$ is proportional to $V_0$ multiplied by the specimen thickness $t$, as expressed in the following equation:

$$\varphi(x, y) = C_E V_0(x, y) t(x, y), \qquad (1)$$

where the incident electron is parallel to $z$ axis of an $x$-$y$-$z$ coordinate system. The parameter $C_E = 0.006526$ $V^{-1}$ $nm^{-1}$ is the interaction constant for an acceleration voltage of 300 kV. To explain the results shown in Fig. 5, $V_0$ is assumed to be constant along the $z$ axis and variable in the $x$-$y$ plane.

Figures 6(a) and 6(b) show a TEM image and a reconstructed phase image acquired from the same area as that of Fig. 5(a). Figure 6c plots the phase shift along the line X1–X2 in Fig. 6(b). Because of the limited distance in the thin-foil specimen, the thickness $t$ can be approximated to be constant over the line X1–X2. Although the phase shift remains almost constant over the region with regular stacking number $n$ = 2 (i.e., $\varphi \approx 2.56$ rad relative to the phase in the reference electron wave), the phase shift changes at the positions indicated by the arrowheads. The magnitude of the additional phase shift is 0.25 rad at the red arrowhead [corresponding to the area shown in Fig. 5(d)], and 0.065 rad at the blue arrowhead [corresponding to the area shown in Fig. 5(c)]. Note that the additional phase shift (0.25 rad and 0.065 rad) does not change appreciably upon tilting the specimen about the $x$ axis from $-3°$ to $+9°$ with respect to the $y$ axis. Thus, diffraction appears to have little effect on the additional phase shift.

To explain the results for the additional phase shift, we calculate the mean inner potential $V_0$, which depends on the irregularity of the stacking number $n$ [i.e., volume fraction of the rock-salt blocks in the extended unit cells], as demonstrated later. The mean inner potential can be approximated by[19]

$$V_0 = \frac{h^2}{2\pi m_0 e \Omega} \sum_j f_j(0), \qquad (2)$$

where $h$ is Plank's constant, $m_0$ is the nonrelativistic electron mass, $e$ is the elementary charge, $\Omega$ is the unit-cell volume, and $f_j(0)$ is the atomic scattering amplitude for constituent element $j$ at the scattering angle $0°$. The summation runs over all atoms in the unit cell. We use the atomic scattering amplitudes of Ca, Sr, Ti, and O calculated by Doyle and Turner using the Hartree–Fock atomic wave functions: $f_{Ca}(0) = 0.9923$ nm, $f_{Sr}(0) = 1.3109$ nm, $f_{Ti}(0) = 0.8776$ nm, and $f_O(0) = 0.1983$ nm.[19,21] For the regions shown in Figs. 5(b)–5(d), we assumed three types of unit cells, as indicated by the red rectangles. Table 1 summarizes the lattice parameters and the number of atoms in these unit cells. To simplify the calculation of the mean inner potential, we assume that the alkaline-earth-metal sites in the perovskite blocks are occupied by Sr ions.

The calculated mean inner potentials are 21.1, 21.9, and 22.9 V for the unit cells shown in Figs. 5(b)–5(d), respectively. Note that the mean inner potential increases with increasing irregularity in the stacking parameter $n$, which is consistent with experiment. Conceptually, the increase in mean inner potential may be explained by an increase in atomic density (due to the reduced fraction of the rock-salt blocks and the increase in Sr content) per unit volume. For further discussion, we also calculate the following two parameters: (1) the mean inner potential of the unit cell in Fig. 5(c) [or Fig. 5(d)] divided by that of Fig. 5(b), which we denote $R_m$, and (2) the phase shift observed in the region of Fig. 5(c) [or Fig. 5(d)] divided by that of Fig. 5(b), which we denote $R_p$. For the region of Fig. 5(c), $R_m = 1.04$ is reasonably consistent with $R_p = 1.03$, and the same can be said for $R_m = 1.09$ and $R_p = 1.09$ in the region of Fig. 5(d). These results show that the effective mean inner potential changes because of the irregularity in stacking number $n$, which explains well the contrast anomaly (i.e., the presence of bright bands) in the HAADF-STEM image.

We briefly discuss the relation between the ferroelectric domain walls and the amount of Sr substitution. As indicated herein, an increase in the Sr substitution should increase the density of the irregular stacking regions composed of the Sr-occupied perovskite blocks. These irregular regions provide preferential sites for the 180° ferroelectric domain walls (Fig. 4). The plate width of the ferroelectric domains is not constant but different by positions because of this relationship. However, our electron microscopy study found another type of irregular regions that were free from the ferroelectric domain walls. The presence of the latter type appeared to result from the excess amount of Sr ions for the formation of the ferroelectric domain walls.

## 4. Summary

The TEM studies reveal the microstructural characteristics of CDWs in the ferroelectric compound CSTO. A systematic series of dark-field images indicates that the morphology of CDWs depends on the viewing direction; in other words, a meandering CDW is observed when viewing along the [001] axis, whereas only short CDW fragments are observed when viewing along the [010] axis because of interception by the 180° ferroelectric domain walls. CDWs from the [001] axis occur regardless of APBs and twin boundaries because there is no definite correlation between those interfaces. The HAADF-STEM observations demonstrated the irregularity of stacking in the perovskite blocks, which appears to affect the stability of the ferroelectric domain walls and CDWs in CSTO. The regions of the irregular stacking showed changes in the mean inner potential due to the Sr content, as indicated by electron holography. These microscopy studies thus improve our understanding of the mechanisms by which ferroelectric domains develop in CSTO.

### Acknowledgments

We thank Dr. Kurushima (Toray Research Center) for the specimen preparation. This study was supported in part by JST CREST (JPMJCR1664) and a Prof. Osafune memorial scholarship by the Japanese Society of Microscopy.

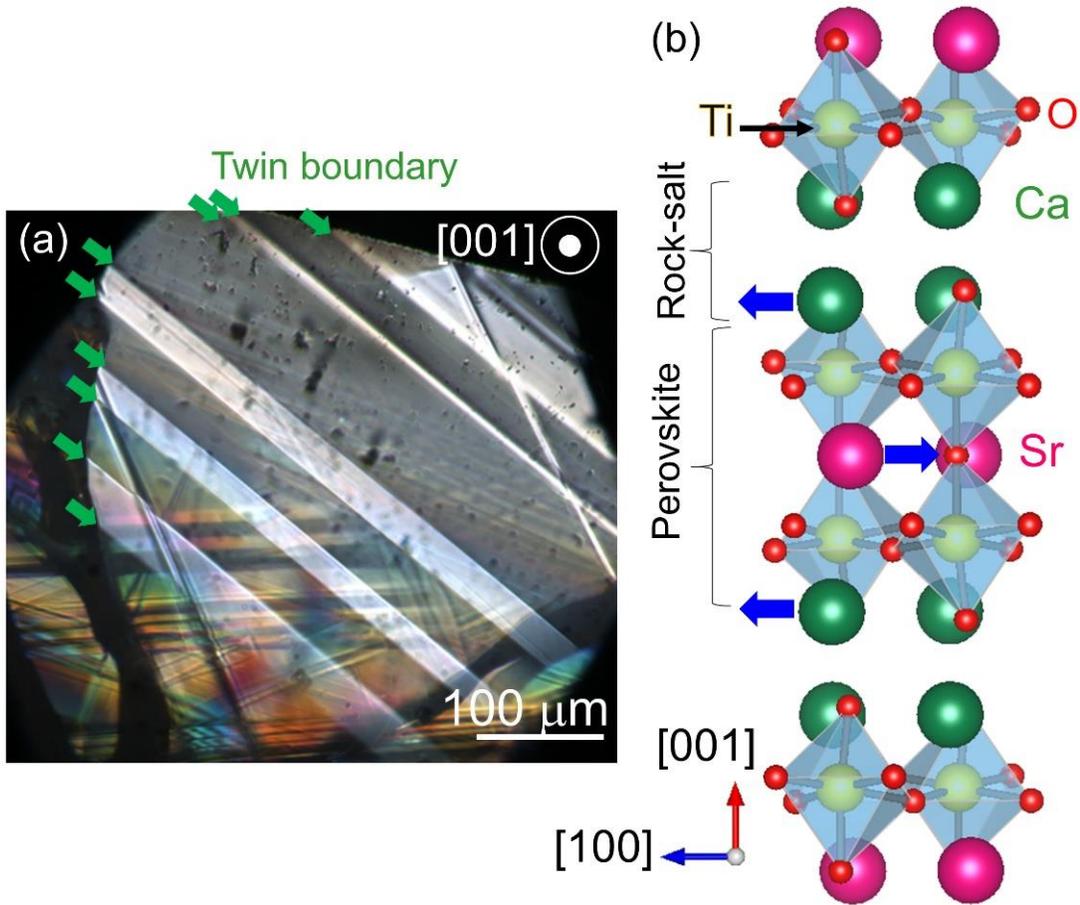

Fig. 1. (a) Optical micrograph of (001) plane. Twin boundaries are indicated with the arrowheads. (b) Crystal structure of $Ca_{3-x}Sr_xTi_2O_7$. The blue arrows represent the displacements of Ca and Sr ions. In the schematic, Sr and Ca ions occupy the alkali-earth-metal sites in the perovskite and rock-salt blocks, respectively.

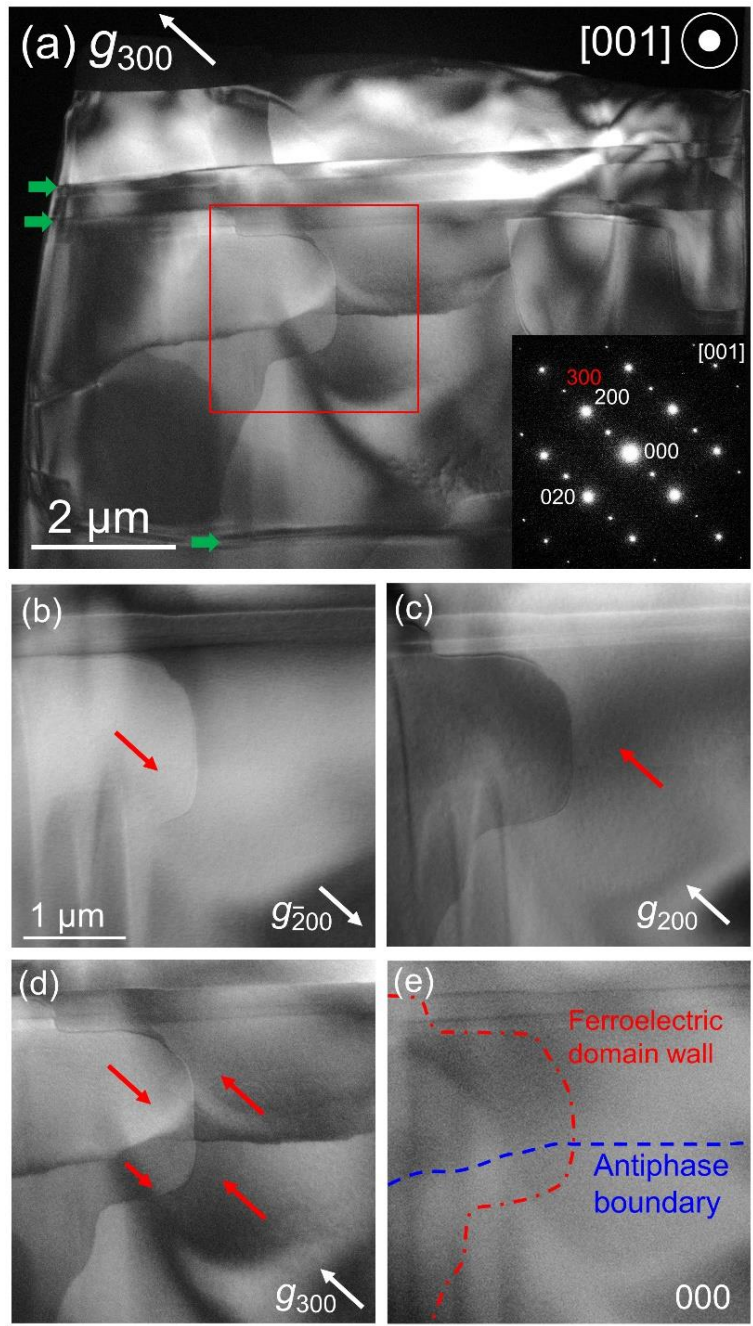

Fig. 2. (a) Dark-field image using the reflection 300 in $Ca_{3-x}Sr_xTi_2O_7$ ($x = 0.45$). The inset is an electron diffraction pattern. The region marked by the red square was observed using the reflections (b) $\bar{2}00$, (c) 200, (d) 300, and (e) direct spot 000. The dark-field images were obtained under the two-beam conditions by tilting the specimen. The red arrows indicate the electric polarization directions at each domain. Panel (e) shows the positions of the ferroelectric domain wall and the antiphase boundary. The green arrows indicate the twin boundaries in panel (a).

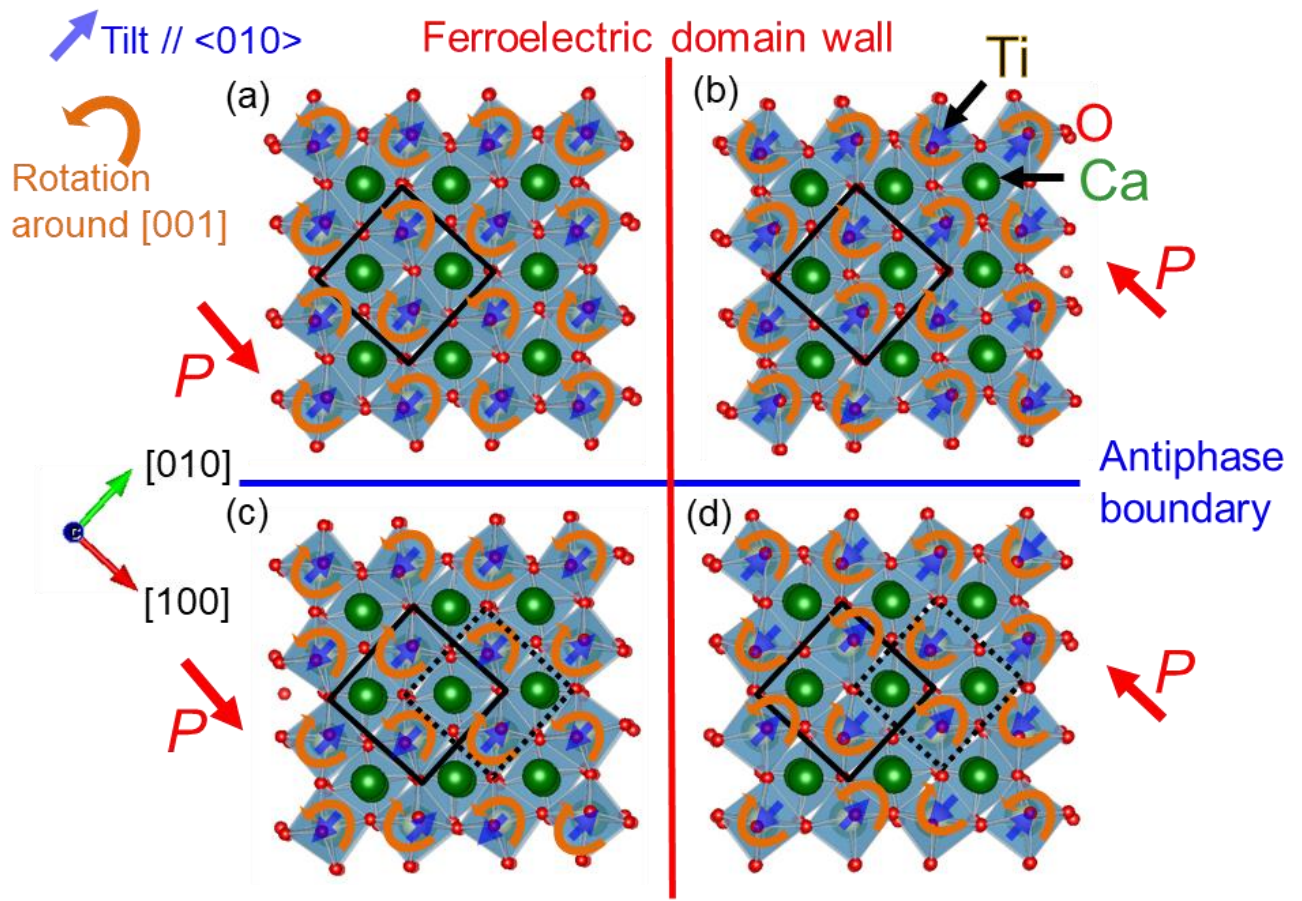

Fig. 3. Schematic illustration of four types of domains resulting from rotation and tilting in Ca$_{3-x}$Sr$_x$Ti$_2$O$_7$. Blue and orange arrows show the tilting along [010] and rotation around [001]. The arrows for the tilting show the direction of displacement of the apical oxygen of the TiO$_6$ octahedron. These symbols are based on a previous report[10]. The electric polarization $P$ is indicated by red arrows in each domain. The squares represent the unit cell of each structure.

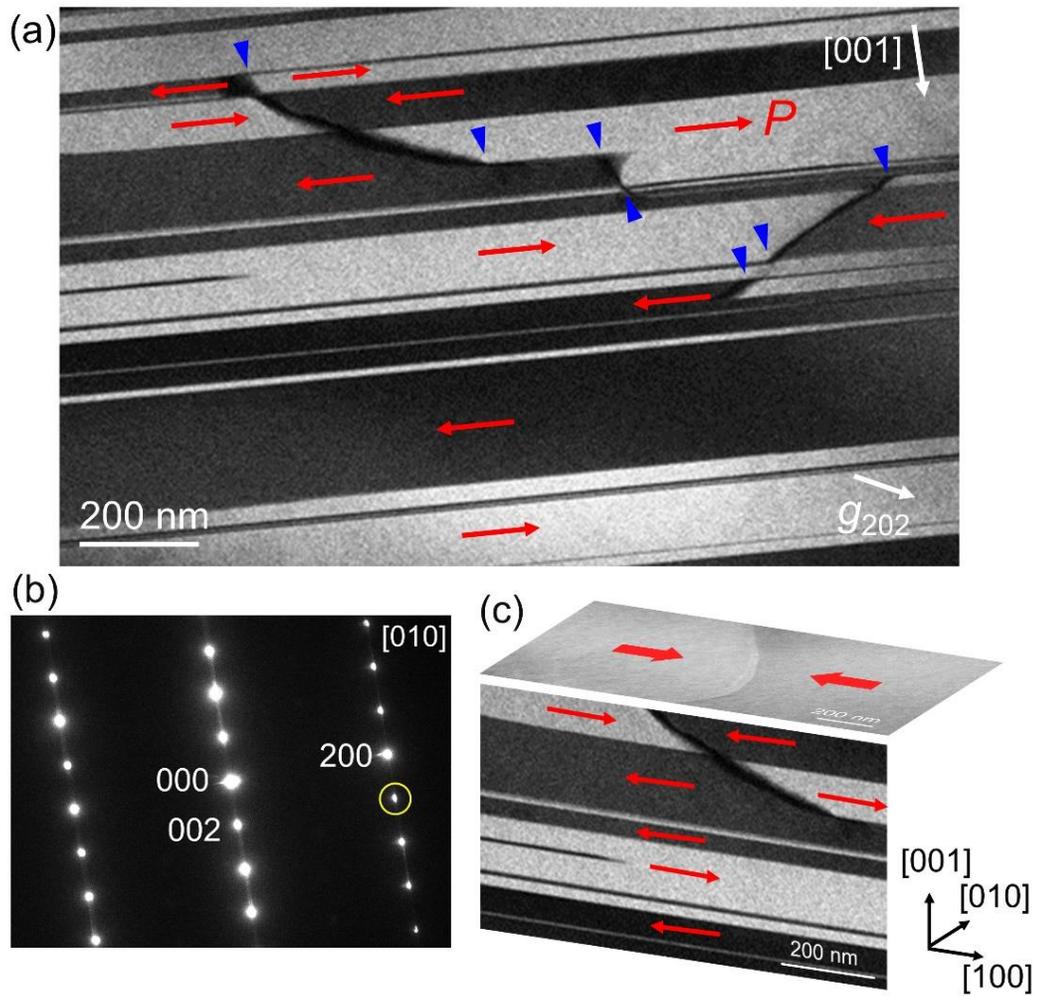

Fig. 4. (a) Dark-field image of reflection 202 in $Ca_{3-x}Sr_xTi_2O_7$ ($x = 0.45$). The electric polarization $P$ of the bright domains is to the right. The blue arrowheads show where CDWs terminate. (b) Selected-area electron diffraction pattern along [010] axis. The yellow circle shows the reflection 202 used in panel (a). Although panel (b) show a diffraction pattern for [010] incidence, the dark-field image (a) was obtained under the two-beam condition, in which the 202 spot is strongly excited. (c) Dark-field images showing top and side views of CDWs.

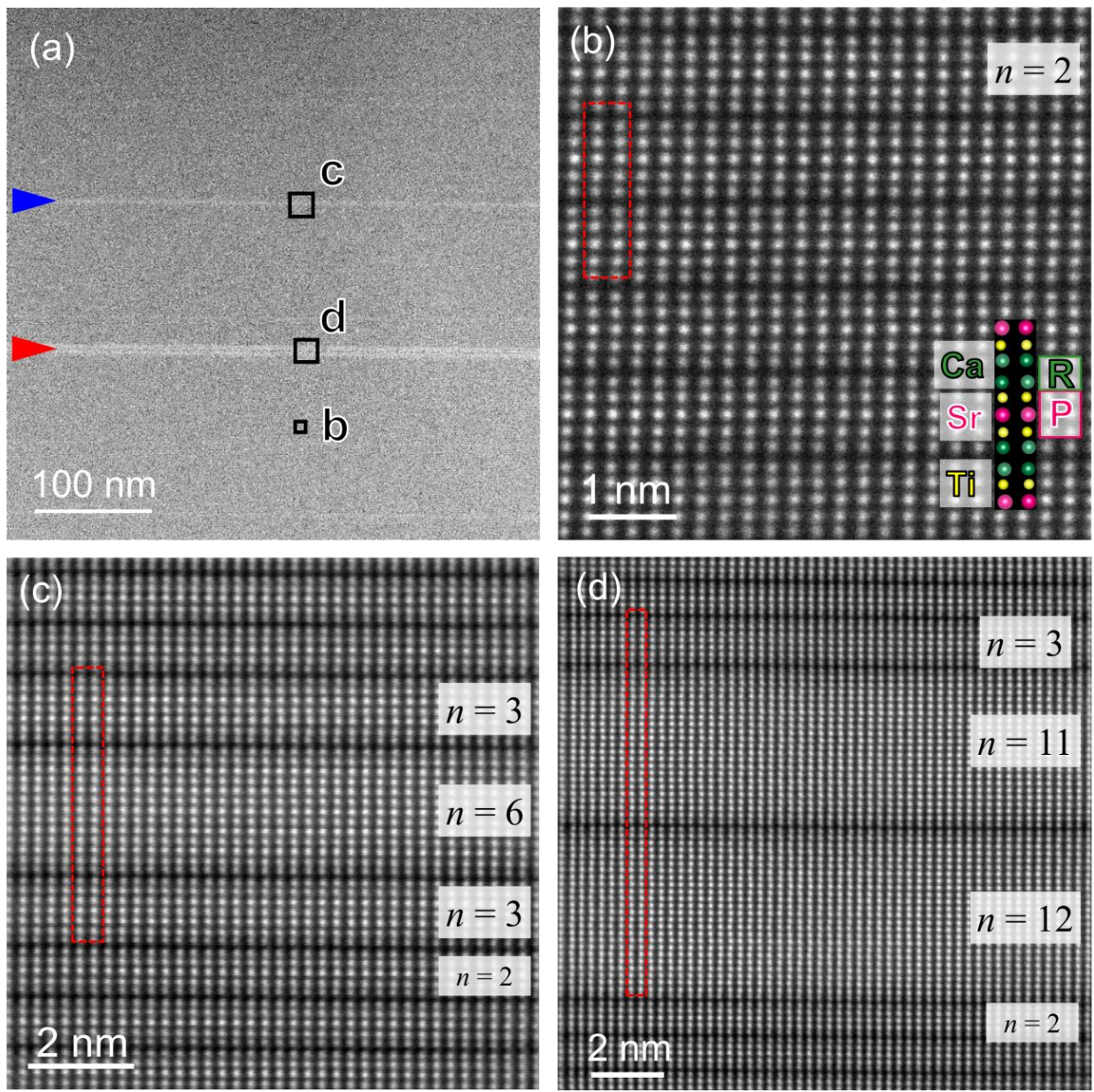

Fig. 5. High-angle annular dark-field scanning transmission electron microscopy images in the 010 plane in $Ca_{3-x}Sr_xTi_2O_7$ ($x = 0.45$). The high-magnification images in panels (b)–(d) were captured from the marked areas. In panel (b), a schematic of the crystal structure is superimposed on the image, with the perovskite (P) and rock-salt (R) blocks marked. Each red rectangle is a lattice (unit cell) for calculating the mean inner potential. The integer $n$ is the number of perovskite blocks.

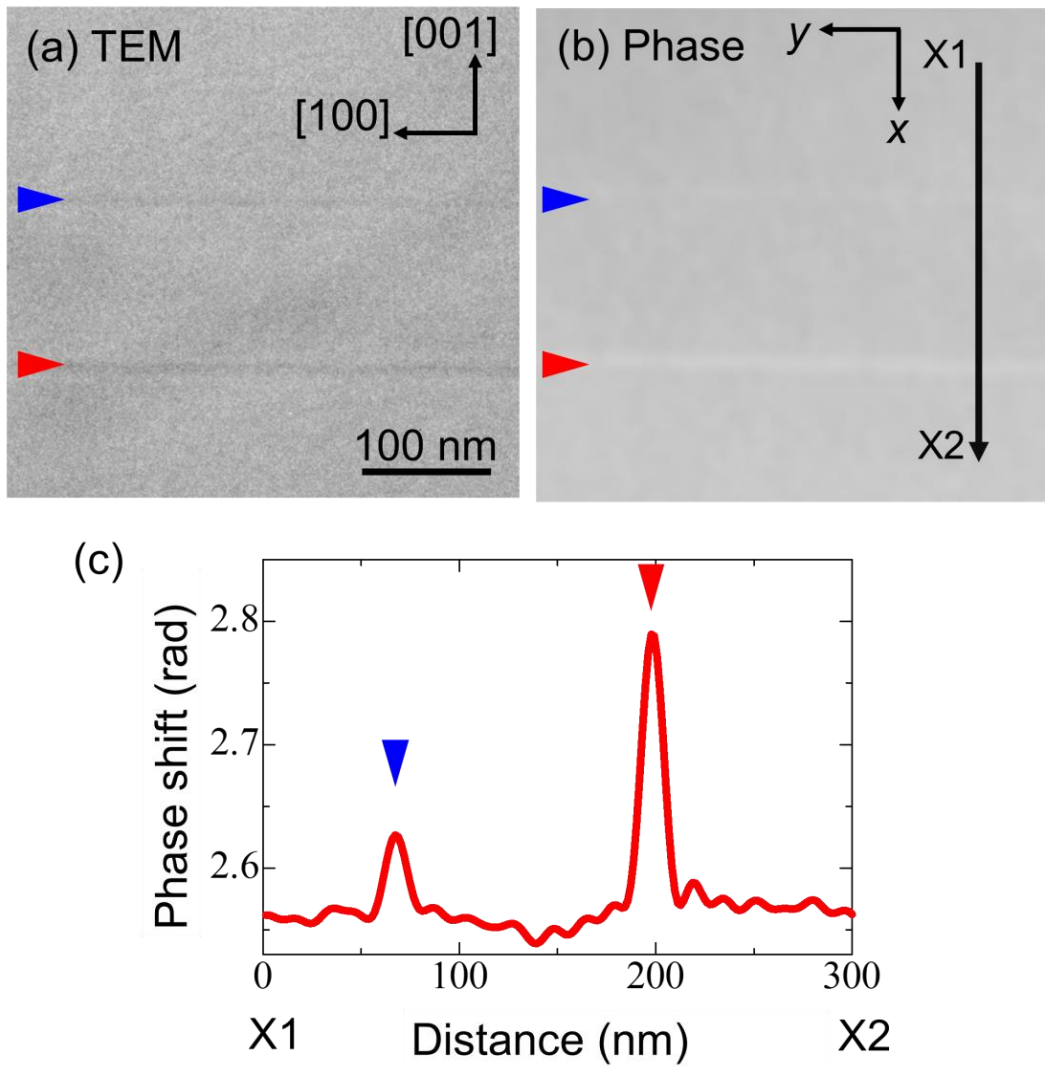

Fig. 6. (a) Transmission electron microscopy image and (b) phase image around the irregular perovskite blocks. The blue and red arrowheads show the same positions as in Fig. 5. (c) Phase shift profile along line X1–X2 in panel (b). The accuracy of the phase shift is approximately 0.016 rad.

**Table 1.** Lattice constant of $c$ axis and number of atoms in each structure of Fig. 5 for calculating the mean inner potential. The calculation uses the lattice constants $a = 5.485$ Å and $b = 5.470$ Å.

|  | $c$ (Å) | Ca | Sr | Ti | O |
|---|---|---|---|---|---|
| Unit cell in Fig. 5(b) | 19.43 | 8 | 4 | 8 | 28 |
| Unit cell in Fig. 5(c) | 52.59 | 12 | 18 | 24 | 78 |
| Unit cell in Fig. 5(d) | 104.3 | 12 | 46 | 52 | 162 |